\documentstyle[aps,epsfig]{revtex}
\begin{document}
\draft
\twocolumn[\hsize\textwidth\columnwidth\hsize\csname
@twocolumnfalse\endcsname
\title{
Coherent oscillation in a linear quantum system coupled to a thermal bath.}
\author{N. F. Bell$^1$, R. F. Sawyer$^2$ and R. R. Volkas$^1$ }
\address{$^1$ School of Physics, Research Centre for High Energy Physics\\
The University of Melbourne, Victoria 3010 Australia\\
$^2$ Department of Physics, University of California at Santa Barbara \\
Santa Barbara, California 93106\\
(n.bell@physics.unimelb.edu.au, sawyer@vulcan.physics.ucsb.edu, r.volkas@physics.unimelb.edu.au)\\
pacs 3.67a, 5.40a}
\maketitle

\begin{abstract}
We consider the time development of the density matrix for a system coupled to 
a thermal bath, in models that go beyond the standard two-level systems through 
addition of an energy excitation degree of freedom and through the possibility 
of the replacement of the spin algebra by a more complex algebra. We find 
conditions under which increasing the coupling to the bath above a certain level 
decreases the rate of entropy production, and in which the limiting behavior is 
a dissipationless sinusoidal oscillation that could be interpreted as the 
synchronization of many modes of the uncoupled system.
\end{abstract}
\vskip2pc]

\noindent
{\it Introduction}.
We explore some extensions of the problem of a two level system in a thermal bath 
\cite{leggett} and find some new qualitative behavior that can be provided by these extensions.
Two specific examples are the symmetrical 
double well in an environment with temperature on the order of the (single-well) 
level spacing, and the propagation of oscillating 
neutrinos in hot, dense media. In both cases we find behavior in which a weak 
coupling of the system
to the bath induces some rate of entropy gain, but where an increase in the 
coupling beyond a certain value 
 reduces the rate of entropy production, reducing  it to zero in a limiting case.
The limiting behavior can correspond to 
the ``frozen" conditions of ref.\cite{H&S,silbey}, and many other papers, in which 
the implicit continuous measurement of a quantity by the system-bath interaction 
can be said to have frozen that quantity in time. 
However our main focus in the present note is on circumstances that give instead a 
kind of dissipationless 
collective oscillation, which gives the appearance of a 
synchronization of modes that had 
different frequencies in the absence of the coupling to the bath. We stress, however, 
that there appears to be little direct connection of these results 
to the large literature of synchronized behavior in non-linear systems \cite{synchro,pant,synchro2} or of  stochastic resonances
in forced systems \cite{stochastic}.

The system itself has two coordinates, which we 
characterize as ``horizontal" and ``vertical". The horizontal component is the ``spin" coordinate of a two state system, 
or in cases that may be of interest in neutrino physics, a ``flavor" coordinate 
for a 3 
or 4-state system. We index these states with Greek letters. The vertical 
variable 
corresponds to a set of energy levels, $E_i$. The system basis states are 
denoted 
$|E_i,\alpha \rangle$. The dynamics of the system uncoupled to the bath is governed 
by 
the Hamiltonian,
\begin{equation}
H_0^{\rm sys}=\sum _{i,\alpha,\beta}\{E_i\delta_{\alpha,\beta}+\lambda_{\alpha,\beta}(E_i)\}|E_i,\alpha \rangle \langle E_i,\beta|.
\label{1}
\end{equation}
We take $\lambda_{\alpha,\beta}(E_i) \ll E_k$ for all values $i,k$. We  consider a 
density matrix for the system $\rho^{\rm sys}(t)={\rm Tr}_{\rm bath}[\rho^{\rm 
s+b}(t)]$, where  $\rho^{\rm s+b}(t)$ is the complete density matrix of the model. 
At an initial time  t=0 we choose a form diagonal in the vertical 
indices,
\begin{equation}
\rho^{\rm sys}(t=0)=\sum_{i,\alpha,\beta}  \rho_{\alpha,\beta}(E_i,t=0)
|E_i,\alpha\rangle\langle E_i,\beta|.
\label{2}
\end{equation}
For these initial conditions, one of the results of a master equation derivation 
is that 
for times $t \gg (\Delta E)^{-1}$, where $ \Delta E$ is a measure of the (single-well) 
energy spacings, the operator $\rho^{sys}$ remains so nearly diagonal in the 
vertical (energy) indices that we can continue to describe the system by a 
vector in the energy space,  $\rho(E_i,t)_{\alpha,\beta}$, replacing ($t=0$) by ($t$) 
in (\ref{2}). Henceforth we suppress the indices $(\alpha,\beta)$ in the horizontal 
space, in 
which $\rho(E_i,t)$ remains an operator.
For the bath-system coupling, $H_{s-b}$, we take the 
dependence on the horizontal (spin-flavor variables) to be in a factorized form, 
$H_{s-b}=\zeta V$, where $\zeta$ operates on the horizontal indices and $V$ 
depends only on the vertical coordinates and the coordinates of the bath. This 
coupling is also taken to be weak, in a way that 
allows 
the problem to be solved, although it can at the same time be strong in another 
sense, for the realization of the limits described above. 
At the end of this note we show that for this form of coupling the  Bloch equations, 
generalized to include the vertical structure, are of the form,
 \begin{eqnarray}
{d\over dt}\rho(E_i,t)&=&-i [\lambda(E_i),\rho(E_i,t)]+\sum_{E_j}\zeta\, 
\rho(E_j,t)\,\zeta \,\Gamma(E_j,E_i)
\nonumber\\
&-&{1\over 2}\Bigr(\zeta^2\, \rho(E_i,t)+\rho(E_i,t) \,\zeta^2 \Bigr) 
\sum_{E_j}\Gamma(E_i,E_j).
\label{3}
\end{eqnarray}
All of the elements in (\ref{3}) are matrices in the horizontal space except for the 
functions $\Gamma$. These functions obey the relation $\Gamma(E_j,E_i)=\exp[(E_j-
E_i)/T]\Gamma(E_i,E_j)$ in consequence of the thermal equilibrium of the bath. 
For the neutrino application, and with the right identifications,  (\ref{3}) is 
the ``Quantum Kinetic Equation" derived by McKellar and Thomson \cite{MandT}. Also 
Loreti and Balantekin \cite{baha}
have used a special case of (\ref{3}) to discuss the energy conserving case of 
propagation of neutrinos through a medium with
a  (spatially) fluctuating electron density, in which case the collision term becomes 
a pure double commutator 
in the horizontal (flavor) space. For the double well application, we sketch 
at the end of this note a derivation in the context needed for our present results.

\underline{ A. Symmetrical double well}

The horizontal space is 2x2, with $\sigma_3=1$ for the particle to be on 
the left side and  $\sigma_3=-1$ for the right side. The barrier height is large 
compared to $\Delta E$ where $\Delta E$ is a 
measure of the energy spacing for the single well.
The vertical states are the single well energy excitations.  We choose 
$\lambda_{\alpha,\beta}(E_i)=[\sigma_1]_{\alpha,\beta}g(E_i)$. 
The energy splittings $g(E_i)$, which are essentially tunneling rates, have strong 
$E_i$ dependence.
For the ``spin" dependence of the coupling to the bath we take, $\zeta=1+b \sigma_3$.
In the case $b=0$ the transition matrix 
elements of 
the operator $V$ for $E_i\rightarrow E_j$ in the left-hand well, considered by 
itself, are the same as those for the right-hand well, as would be the case in the 
dipole approximation of the interaction with a radiation field. For the case $b=1$
 only the amplitude on the left-hand side of the well interacts with the bath.

The coefficient $\Gamma(E_i,E_j)$ is the (single well) 
rate 
for a state $E_i$ to make a transition to state $E_j$.  
When $T\approx \Delta E$, where $\Delta E$ is the scale of the 
vertical splittings, the bath induces transitions among the vertical 
states, and in this case our 
models differ substantively from models that do 
not 
have the vertical structure. 

\underline{ B. Neutrinos}

The question is the time dependence of the flavor density matrix for neutrinos 
that 
oscillate and are also scattered as they traverse a dense medium, like that of 
the 
supernova core, or the early universe at a temperature greater than 1 MeV. Here 
the 
$\lambda_{\alpha,\beta}$ parameters are the combination
of the mass$^2/(2E)$ matrix and the ``index of refraction" terms, the latter 
energy 
independent and of order $G_W$. The functions, $\Gamma (E_i,E_j)$, of order 
$G_W^2$, are the rates of transition for a neutrino to go from state
$E_i$ to state $E_j$, as calculated with the matrix $\zeta$ set equal to unity. 
The 
choice $b=0$ would apply to $\nu_\mu, \nu_\tau$ mixing  in a flavor-blind medium, 
that is, a medium in which 
$\nu_\mu$ and $\nu_\tau$ have the same interactions with the environment.
The choice $b=1$ corresponds to ``active-sterile" scenarios in 
which 
one of the neutrinos does not interact with the environment at all.

\medskip
\noindent
{\it Results for $b=0$}. 
As an example we use a double square well with infinite barriers to the left and right 
of $x=\pm 8a$, and with a
central barrier of width $2a$ and height $U_0$. We choose  $U_0$ and the particle 
mass such that
twenty states are bound with energies less than $U_0$. The 
low-lying states are thus very nearly the symmetrical and anti-symmetrical 
combinations of 
separate states in an infinite well of width $7a$. The energy splittings, $g(E_i)$ 
are easy to calculate 
in the limit of small tunneling.  For the bath we take bosons with $\omega=vk$, 
equal energy spacing
 $2\pi v/L$ where $L\rightarrow \infty$, and with creation and annihilation 
operators, $a_n ^\dagger , 
a_n$. We take $H_0=\sum_n a_n ^\dagger a_n \omega_n+H_0^{({\rm sys})}$, and in the 
system-bath interaction, $H_{s-b}=V\zeta$, we take the vertical operator $V$ to 
have off-diagonal matrix elements given in dipole expansion as
\footnote
{There are diagonal matrix elements of the $kx$ operator as well, since the wells 
are not centered at $x=0$. In principle,
diagonal matrix elements contribute to (\ref{3}) in linear order, as a modification 
of the oscillation matrices $\lambda$, but in the present
case these contributions cancel when $-k$ is added to $k$.},
\begin{equation}
V_{j,k}=
 {q\over L}\sum_n[(a_n ^\dagger +a_n )\delta_{j,k}+ik_nx_{j,k}(a_n ^\dagger -a_n )].
\label{4}
\end{equation}
where $q$ is a coupling strength.
We calculate the off-diagonal dipole matrix elements, $x_{j,k}$ using the infinite 
well wave-functions and define $c_{j,k}=qk_n x_{j,k}$, 
where the later formalism will justify using (near) energy conservation to 
express $k_n$ as $\pm(E_j-E_k)/v$. The coefficients in the bath interaction part of (\ref{3}) are then,

\begin{equation}
\Gamma(E_j,E_k) =2 \pi c_{j,k}c_{k,j}\Bigr [{ \theta(E_j-E_k)
\over e^{(E_j-E_k)/T}-1}+ {\theta(E_k-E_j) \over 1-e^{(E_j-E_k)/T}}\Bigr].
\label{11}
\end{equation}

Incorporating these results and solving (\ref{3}) for several values of $q$, we show in Fig.1 the probability that the particle 
will be on the left side at time $t$, 
with the initial condition of being on the left and in a thermal distribution at the bath 
temperature, which we take as  $T=5$ in units of the single-well ground state energy.  

\begin {figure}[h]
    \begin{center}
        \epsfxsize 3in
        \begin{tabular}{rc}
            \vbox{\hbox{
$\displaystyle{ \, { } }$
               \hskip -0.1in \null} 
} &
            \epsfbox{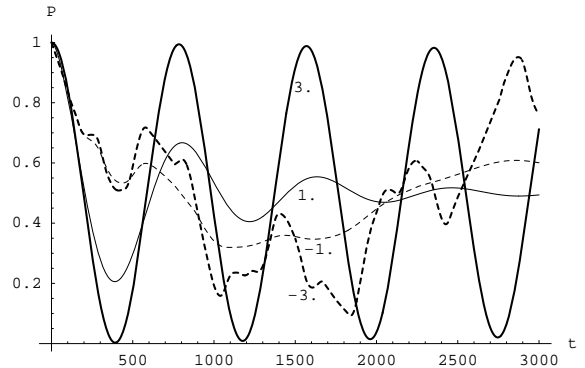} \\
            &
            \hbox{} \\
        \end{tabular}
    \end{center}
\label{fig1}
\protect\caption
	{%
The probability, P, for a particle to be found in the left hand well, for a bath temperature $T=5$, in units of the ground state energy of the single infinite well. The initial condition is that the particle is on the left with a thermal distribution of energies. The integers (-3,-1,1,3) labeling the curves are the Log$_{10}$ of the bath-system coupling strengths in arbitrary units.
	}
\end {figure}

For very strong coupling, the probability becomes simply periodic, as one mode would be 
in the absence of dissipation. This limit can be solved analytically, with the result that the 
oscillation frequency is the thermal average of the oscillation frequencies for the separate modes \cite{bell}.
In fig 2 we show the entropy, as defined by $S=\sum_j{\rm Tr} [\rho(E_j)\log(\rho(E_j))]$, where the trace is over the $2\times2$ space. 
We note that as the coupling is increased from small to moderate values the rate of entropy 
increase rises, but that as the coupling becomes very large the rate of entropy increase goes to zero.

\begin {figure}[h]
    \begin{center}
        \epsfxsize 3in
        \begin{tabular}{rc}
            \vbox{\hbox{
$\displaystyle{ \, {  } } $
               \hskip -0.1in \null} 
} &
            \epsfbox{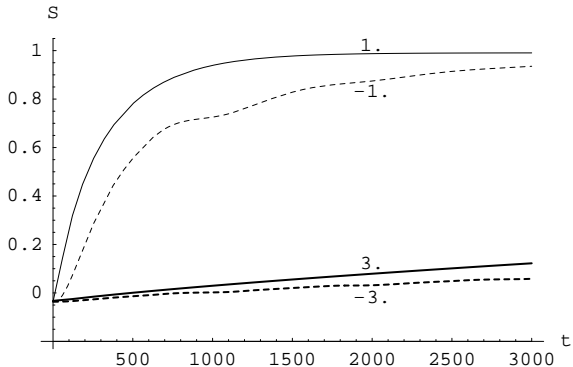} \\
            &
            \hbox{ } \\
        \end{tabular}
    \end{center}
\label{fig2}
\protect\caption
	{%
The entropy S as a function of time for the same values of Log$_{10}$[coupling] used in fig.1.
	}
\end{figure}

We can generalize (\ref{3}) to apply to a system in which the wells contain N fermions with no 
mutual interactions. For example, after making the appropriate modifications to the master 
equation for Fermi statistics, and using the same values of parameters as before, but 
taking $N= 5$, all initially on the left and in thermal equilibrium, we again look at the strong 
coupling limit, getting results very similar to those shown if fig.1 for the density matrix of 
a single particle.
 It is surprising that
one can prepare a state in which, if we looked only at number density,  a set of Fermi particles could appear to be 
oscillating, all in phase, between the left and the right, as would an assemblage of bosons all 
in an ``average energy" state.

For the case of neutrino mixing in a noisy environment, where the vertical space is a continuum, the generalized Bloch equations, (\ref{3})
are integro-differential equations. But we find that 
when they are discretized with, say, around  30 energy levels, they are easily solved, 
with the results insensitive to the mesh size. The results are qualitatively 
the same as those discussed above.

There is an amusing generalization of the above considerations in the extension to a double well with a time dependent bias, 
$\delta H_0^{\rm sys}= \epsilon(t) \sigma_3$. Consider a case in which we begin with $\epsilon>0$ and $\epsilon \gg| g (E_i)|$ for 
some set of low-lying states  (but $\epsilon\ll \Delta E$).
A group of particles initially localized on the left, distributed among 
the low-lying states, and not interacting with the medium, will stay largely on the left  if the bias 
is maintained at the initial value. Now let the bias be slowly lowered until it is equal and opposite to the original bias. 
The particles will efficiently be carried to the right hand well. This is the analogue of the MSW transition 
in neutrino physics. It is a rather diffuse transition, since the different energy levels go 
through their avoided crossings at different times. With a moderate coupling to the bath, the 
 system instead settles down rather quickly to half on the right and half on the left, with no interesting
 persistent order. However, if we increase the bath coupling sufficiently  we regain the shift from left to right. The transition is now sharper than it
 was in the case without the bath. Detailed analysis of this type of behavior for the neutrino case is given in \cite{bell}.

\medskip
\noindent
{\it Results for $b\ne 0$}.
We briefly summarize some examples. In the case plotted in fig. 1, for the strongest coupling, 
we find little change for $b=0.001$.  But $b=0.005$ starts to poison the oscillation, damping it by a factor of 20\%
in 5  periods.  As we move to much larger $b$,  the motion becomes more frozen, so that when, e.g., $b=0.5$,  98\% of 
the probability has remained on the left hand side over a time span of five average periods of the $b=0$ system.  
A popular interpretation  of 
frozen behavior is that the interactions with the bath are continuously measuring the 
value of $\sigma_3$ and projecting the wave function onto the initial state. Although there is no measurement implied by our
equations, the results are consistent with this terminology, 
since in the case with $b=0$ the system-bath interaction does not distinguish the right-hand well from the left-hand 
well and it is for this case that we get the limiting behavior of undamped simple oscillations with a thermal 
averaged frequency.

\medskip
\noindent
{\it Derivation of the master equation}.
 We focus on the aspects that give rise to the placing of the horizontal matrices in (\ref{3})
and the detailed form of the RHS of (\ref{11}). The methods are closely related to those given in \cite{vankampen} and \cite{walls}.
We introduce an interaction picture based on the above division of the Hamiltonian. 
The  
system-bath coupling in this picture is denoted by $H_I(t)$,  with vertical matrix elements,
\begin{equation}
 [ H_I(t)]_{j,k}=e^{i\lambda(E_j) t}\zeta e^{-i\lambda(E_k) 
t}V^I_{j,k}e^{i(E_j-
E_k)t}.
\label{5}
\end{equation}
The leading terms in the system+bath density matrix, $\rho^{\rm 
s+b}_{I}$, in this interaction picture are generated by the integral equation,

\begin{eqnarray}
\rho^{\rm s+b}_I (t) &=& \rho^{\rm s+b}(t=0) \nonumber\\
&-&\int_0^t dt_1\int_0^{t_1}dt_2\Bigr[ 
[\rho^{\rm s+b}_I(t_2),H_I(t_2) ],H_I(t_1)\Bigr ].
\label{6}
\end{eqnarray}

In the perturbation expansion of (\ref{6}) we find a piece with an additional power of the quantities $c_{j,k}^2   [E 
t, E \lambda^{-1}]$ in each higher order. This provides
our definition of leading terms and the explanation  for how a tiny $c_{j,k}^2$ can lead to appreciable effects
over long periods of time. These terms are generated by the terms in which each successive new pair of $H_I$'s in the 
iteration solution puts the state back to the same vertical (and bath) state. 
These observations provide the basis for having omitted, in writing (\ref{6}), all 
 odd terms in $H_I$, and they also dictate the 
form of the density matrix to be used in solving the equation, 
\begin{equation}
\rho^{\rm s+b}_I=\rho^{\rm bath}\sum_j \rho_I(E_j,t) |j \rangle \langle j|.
\label{7}
\end{equation}
where,
\begin{equation}
\rho^{\rm bath}=\Bigr[\prod_i(1-e^{-\omega_i/T})\Bigr]\exp\Bigr [-(\sum_n a_n 
^\dagger a_n \omega_n)/T \Bigr ],
\label{8}
\end{equation}
and where,
\begin{equation}
\rho_I(E_j,t)=e^{i\lambda (E_j)t}\rho(E_j,t)e^{-i\lambda (E_j)t}.
\label{9}
\end{equation}
We substitute  (\ref{7}) into (\ref{6}), take the bath trace, and use leading 
order 
expressions for each term in the double commutator, obtaining, for example,
\begin{eqnarray}
&{\rm Tr}&_{\rm bath}\Bigr[\langle j|H_I(t_1)\rho^{\rm s+b}_I 
(t_2)H_I(t_2)|j\rangle \Bigr] =\delta(t_1-t_2)  \nonumber\\
&\times& \sum _k \Gamma(E_k,E_j) 
e^{i\lambda(E_j) t_1}\zeta e^{-i\lambda(E_k)
t_1} \nonumber\\
&\times& \rho_I(E_k,t_1)e^{i\lambda(E_k) 
t_1}\zeta e^{-i\lambda(E_j) t_1}.
\label{10}
\end{eqnarray}
Similarly, we have
\begin{eqnarray}
&{\rm Tr}&_{\rm bath}\Bigr[\langle j|H_I(t_1)H_I(t_2)\rho_I^{\rm 
s+b}(t_2)|j \rangle \Bigr]=\delta(t_1-t_2) \nonumber\\
&\times& e^{i\lambda(E_j)
t_1}\zeta^2e^{-i\lambda(E_j 
)t_1}\sum _k \Gamma(E_j,E_k) \rho_I(E_j, t_1).
\label{12}
\end{eqnarray}
The key to separation of the leading terms in the above is performing the sum 
over the 
modes of the scalar field first. After making the transition to the continuum, 
we make replacements of the form, 
\begin{eqnarray}
&\int_{\omega_{min}}^\infty & d \omega f(\omega) e^{i(\pm E_j \mp E_k \pm
\omega)(t_1-t_2)} \nonumber\\
&\rightarrow& 
2\pi  
f[(E_k-E_j)]\delta(t_1-t_2)\theta(E_k-E_j).
\label{13}
\end{eqnarray}
where neither the introduced infrared cut-off, $\omega_{min}$ nor the residual terms contribute to leading 
order, as defined above.

Using  (\ref{10}), (\ref{12}) and their counterparts for the two other 
orderings in 
({6}) , doing the $t_2$ integral (with the delta function symmetrically smeared, as can be 
justified by more accurate integrations) and differentiating  with respect to $t$ gives,
\begin{eqnarray}
&{d\over dt}&\rho_I(E_j,t)=\sum_{E_k}e^{i\lambda(E_j) t}\zeta
e^{-i\lambda(E_k) t} 
\rho_I(E_k,t) \nonumber\\
&\times& e^{i\lambda(E_k) t}\zeta e^{-i\lambda(E_j) t}
\Gamma(E_k,E_j)
\nonumber\\
&-&{1\over 2}\Bigr(e^{i\lambda(E_j) t}\zeta^2e^{-i\lambda(E_j )t} 
\rho_I(E_j,t) \nonumber\\
&+&\rho_I(E_j,t) e^{i\lambda(E_j)
t}\zeta^2e^{-i\lambda(E_j )t} 
\Bigr) 
\sum_{E_k}\Gamma(E_j,E_k).
\label{14}
\end{eqnarray}
Using (\ref{9}) we now 
regain (\ref{3}).

To summarize, we have developed the elements of the master equation for a bath-coupled
system that is a generalization in two directions of the system of the ground states of a double-well.
On the basis of this equation we have found new instances in which strong bath-system interactions
can lead to more, rather than less, orderly behavior. We have not begun to address the 
question of identifying a realistic quantum well system
in which one could make an experimental application
of the conclusions. We should emphasize, however,
that our limit of ``large coupling" to the bath is the limit
in which $\Gamma \gg \lambda$; it is achievable, in principle,
for any strength of the system-bath coupling (whether the mechanism be photons,
phonons or collisions with surrounding particles), simply by widening the
barrier between the wells.

We thank Wick Haxton for hospitality at the National Institute for Nuclear Theory, where conversations among the three 
us began. The work of R.F.S. was supported, in part, by NSF grant PHY-9900544,
N.F.B. by the Commonwealth of Australia, and R.R.V. by the Australian Research
Council. R.F.S. would like to thank B. McKellar for kind hospitality during a visit
to the University of Melbourne. R.F.S. thanks Horia Metiu for an enlightening
conversation.

\end{document}